\documentclass[12pt,letter]{iopart}
\usepackage{amssymb}
\usepackage{graphicx}

\newcommand{\as}         {$^{75}$As}

\newcommand{\lsxco}     {${\rm La}_{2-x} {\rm Sr}_{x} {\rm Cu O_4}$}

\newcommand{\slrr}      {$T_1^{-1}$}

\newcommand{\slrrtext}  {spin lattice relaxation rate}

\newcommand{\cafeas}    {CaFe$_2$As$_2$}
\newcommand{\fpn}    {ferropnictides}

\begin{document}

\title[NMR in CaFe$_2$As$_2$]{Low energy spin dynamics in the antiferromagnetic phase of CaFe$_2$As$_2$}

\author{N. J. Curro, A. P. Dioguardi, N. ApRoberts-Warren, A. C. Shockley and P. Klavins}

\address{Department of Physics,
University of California, One Shields Avenue, Davis, CA 95616-8677}
\ead{curro@physics.ucdavis.edu}

\begin{abstract}
We present \as\ nuclear magnetic resonance data in the paramagnetic and magnetic states of single crystal \cafeas.  The electric field gradient and the internal magnetic field at the As sites change discontinuously below the first order structural transition at $T_0=169$ K. In the magnetic state, we find a single value of the internal hyperfine field consistent with commensurate antiferromagnetic order of Fe moments pointing in the $ab$ plane.  The spin lattice relaxation rate shows Korringa behavior for $T \lesssim T_0/3$, reflecting the metallic nature of the ordered state. Surprisingly,  \slrr\ exhibits a small peak at 10 K, revealing the presence of slow spin fluctuations that may be associated with domain wall motion.
\end{abstract}

\submitto{\NJP}
\maketitle

\section*{Introduction}

The discovery of the ferropnictide superconductors has generated tremendous interest in the scientific community because of their high transition temperature and potential for applications.  One of the central questions surrounding these materials is the nature of the superconductivity and whether or not it is connected to the antiferromagnetism present in this family of compounds.  There is compelling evidence to suggest that the ferropnictides have properties similar to a number of strongly correlated electron materials in which unconventional superconductivity emerges in proximity to antiferromagnetic ground states \cite{BuchnerNature2009,uedareview}.  Yet there are a number of important distinctions, foremost of which is that the antiferromagnetic parent state of the \fpn\ is an itinerant spin density wave, whereas that of the cuprates is Mott insulating.  It is not clear whether the magnetic excitations of these two very different states could be responsible for superconductivity, or whether an unidentified novel pairing mechanism is at play in the ferropnictides \cite{MonthouxPinesReview,MazinPairingPRL2008}.  Therefore, it is crucial to characterize the ordered antiferromagnetic state of the \fpn\ to fully understand the nature of the excitations of the parent material.

In order to address these questions we have conducted detailed NMR studies in the antiferromagnetic phase of \cafeas.  We have measured the internal hyperfine field, electric field gradient (EFG) and spin lattice relaxation to investigate the low energy spin dynamics of the ordered state.  \cafeas\ is an ideal system for NMR investigations, as clean single crystals are easily grown, and the EFG is sufficiently large that spectra can be obtained without an externally applied magnetic field.  In the antiferromagnetic state, the ordered Fe moments give rise to a static internal magnetic field at the As sites that is directly proportional to the sublattice magnetization. This static field is aligned along the $c$-direction and is single-valued, consistent with commensurate stripe-like magnetic order as detected by neutron scattering \cite{Goldman2008PRB}.  We find that the order parameter varies discontinuously at the structural-magnetic transition at  $T_0 = 169$ K.  Below this temperature we find that the nuclear \slrrtext, \slrr, is strongly temperature dependent.  Our data suggest that the nuclear spins are relaxed both by collective magnetic excitations as well as itinerant quasiparticle scattering. Surprisingly, however, we find that \slrr\ increases below  $\sim 20$ K, reaching a peak at 10 K before falling below this temperature.  This low temperature peak reveals the presence of slow spin fluctuations, and is reminiscent of the glassy behavior found in the lightly doped cuprates \cite{currohammel}.   These slow spin fluctuations may be related to the motion of antiferromagnetic domain walls that has been discussed recently as a explanation for the reduced ordered moment in these materials \cite{MazinNature2009}.

The magnetic state of the ferropnictides is unusual, consisting of columnar antiferromagnetic arrangements of frustrated Fe spins with reduced ordered moments.  Electronic structure calculations capture many of the salient features of the Fermi surface and magnetic structure, but predict an ordered moment on the order of 2$\mu_B$ \cite{YinPickett, Hirschfeld2008}. Conversely, experiments reveal ordered moments on the order of 0.3-0.8$\mu_B$ \cite{Cruz2008,Goldman2008PRB,HenningPRL2008,takigawa2008}.  There are three points of view that have emerged in the literature concerning this discrepancy: (i) the Fe moments are reduced by spin-orbit and hybridization effects \cite{ZaanenOrbitalSpinPRB2009,CastroNetoReducedMoment}, (ii) a complex domain structure in the ordered state with mobile fluctuating domain walls that reduce the effective moment \cite{MazinNature2009}, and (iii)  proximity to a quantum phase transition reduces the ordered moment \cite{Si2008,Uhrig2008}. Regardless of the particular scenario, the nature of columnar antiferromagnetic order naturally provides for several types of domain walls.  Since motion of the domain walls typically involves collectively flipping several spins simultaneously, the time scale for this motion can reach $10^{-6}$ sec, on the order of the NMR Larmor frequency, and the gradual slow-down of such fluctuations can give rise to a Bloembergen-Purcell-Pound (BPP) peak in the NMR relaxation rate \cite{BPP}.  We are able to fit the \slrr\ data with an activated spin-flip model, where the activation energy $\Delta\sim 5$ K.

\section*{Synthesis and Characterization}

\begin{figure}
\begin{center}
 \includegraphics[width=120mm,clip]{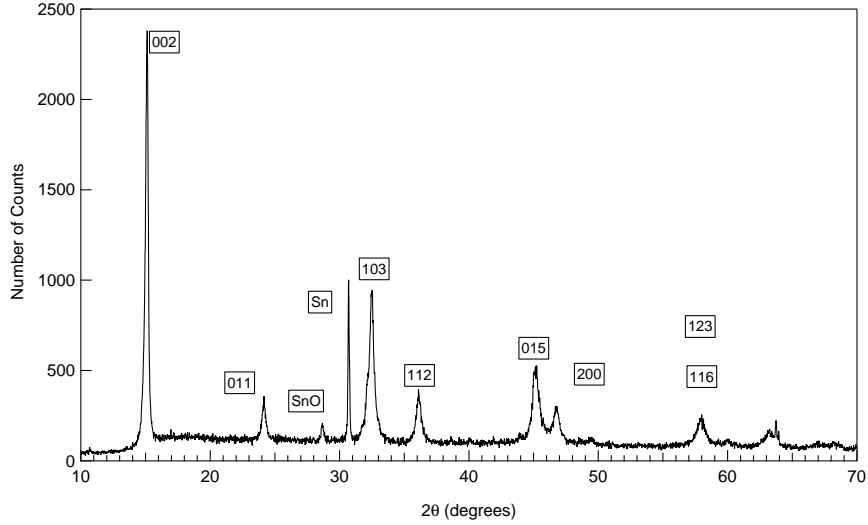}
 \caption{The powder x-ray diffraction spectrum of \cafeas, with individual peaks labeled.  The peak labeled Sn corresponds to Sn flux.}
\label{fig:xray}
\end{center}
\end{figure}

Large single crystals of \cafeas\ were grown in Sn flux in the stoichiometric ratio Ca:Fe:As:Sn = 1:2:2:20 via a standard method which is discussed in \cite{RonningCaFe2As2discovery}.   The elemental materials were placed in a 10mL alumina crucible and sealed in an evacuated quartz ampoule.  The ampoule was heated to 500 C at 125 C/hr with a dwell time of 6 hours.  The temperature was then increased at 100 C/hr to 750 C, 950 C, and eventually 1100 C and held for 8, 12, and 4 hours respectively.  After this, the temperature was decreased at a rate of 4 C/hr to 600 C,  the ampoule was removed from the furnace, and the Sn flux was removed with a centrifuge.
We note that two distinct crystal morphologies were produced.  Single crystals of \cafeas\ were present in plate-like form with dimensions on the order of 3 x 3 x 0.5 mm.  However, a second phase of needle-shaped crystals also formed which remains uncharacterized.  The two phases grew on top of each other, and as a result, the yield of unadulterated \cafeas\ was significantly diminished.

\begin{figure}
\begin{center}
 \includegraphics[width=75mm]{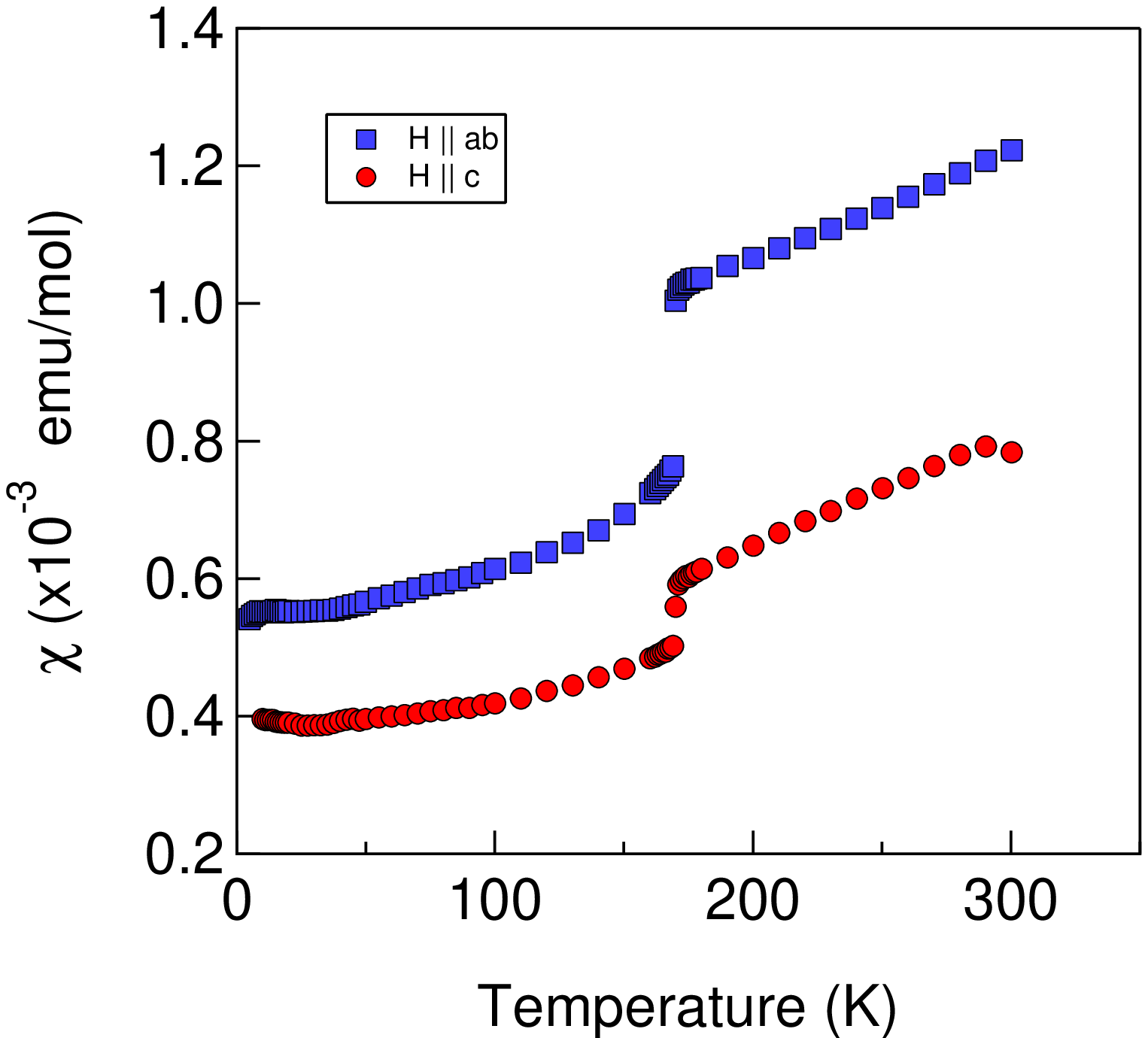}
 \includegraphics[width=80mm]{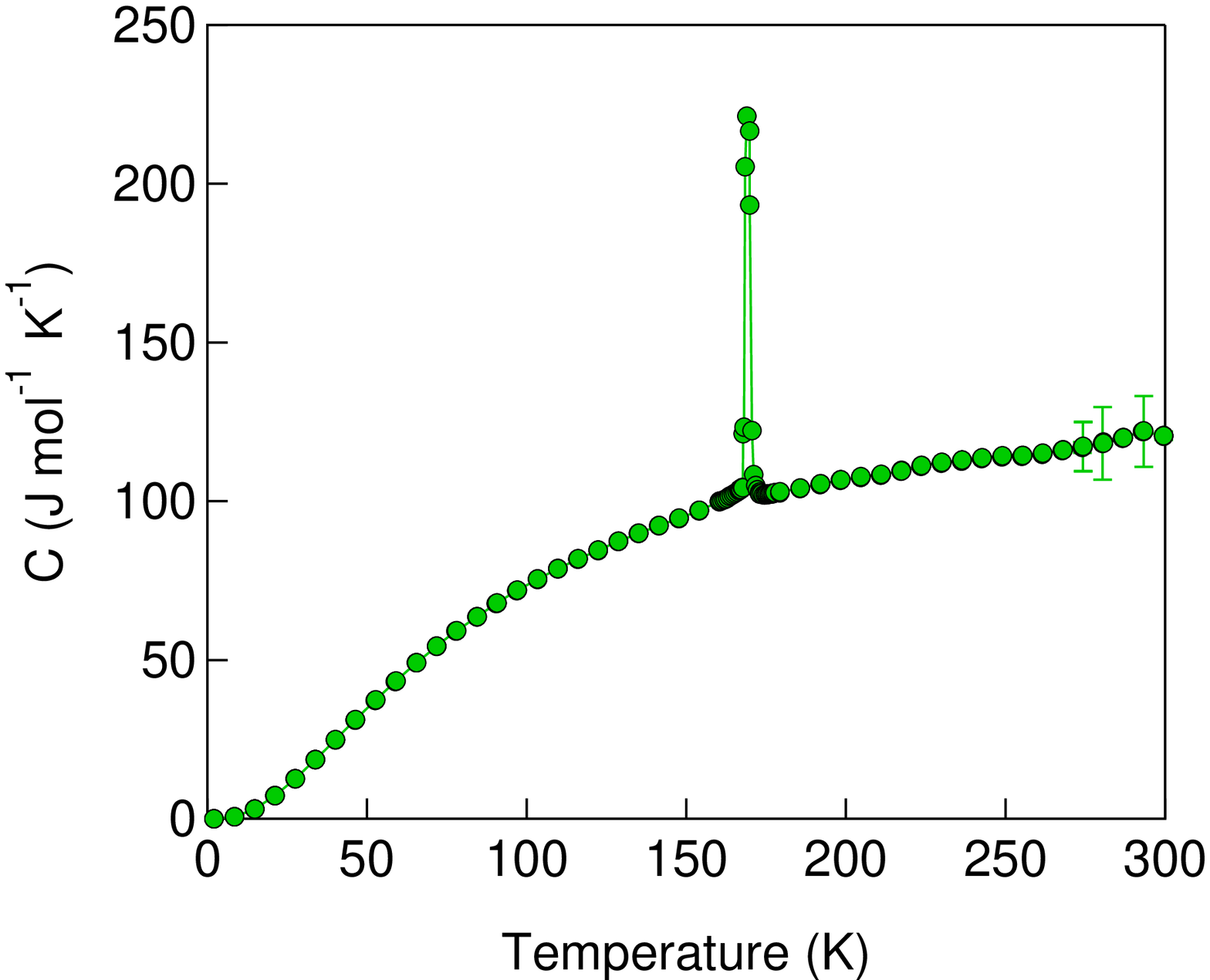}
\caption{(Left) The magnetic susceptibility along the $c$ and $ab$ directions in single crystal \cafeas, as measured in a field of 10 Oe. (Right) The heat capacity of \cafeas\ as a function of temperature. The first order transition is clearly visible in both sets of data at $T_0 = 169$ K.}
\label{fig:chiandC}
\end{center}
\end{figure}

Powder diffraction (Fig. \ref{fig:xray}) of the plate-like \cafeas\ materials revealed the $I4/mmm$ tetragonal structure consistent with previously reported data. Magnetic susceptibility as well as specific heat (Fig. \ref{fig:chiandC}) show a clear first order transition at $T_0 = 169$ K  \cite{RonningCaFe2As2discovery}.  Furthermore, the Nuclear Quadrupolar Resonance (NQR) spectrum remains sharp (FWHM $\sim$ 480 kHz) and in quantitative agreement with previous data \cite{BaekCaFe2As2PRB}.  We are therefore  confident that our measurements probe the intrinsic behavior of the \cafeas\ phase.

\section*{Magnetic Resonance}

\subsection*{Spectral Measurements}
\begin{figure}
\begin{center}
 \includegraphics[width=70mm]{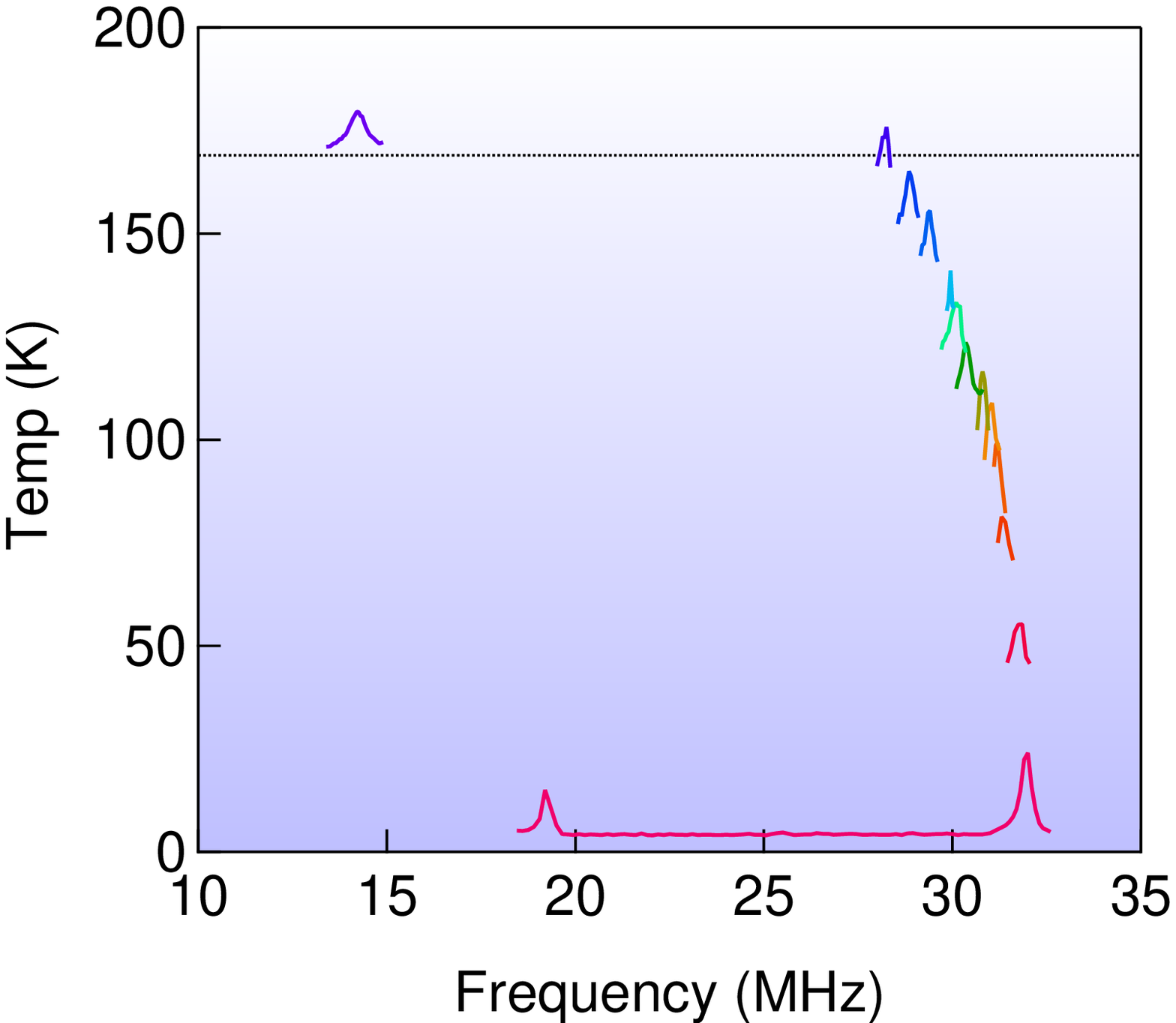}
 \includegraphics[width=60mm]{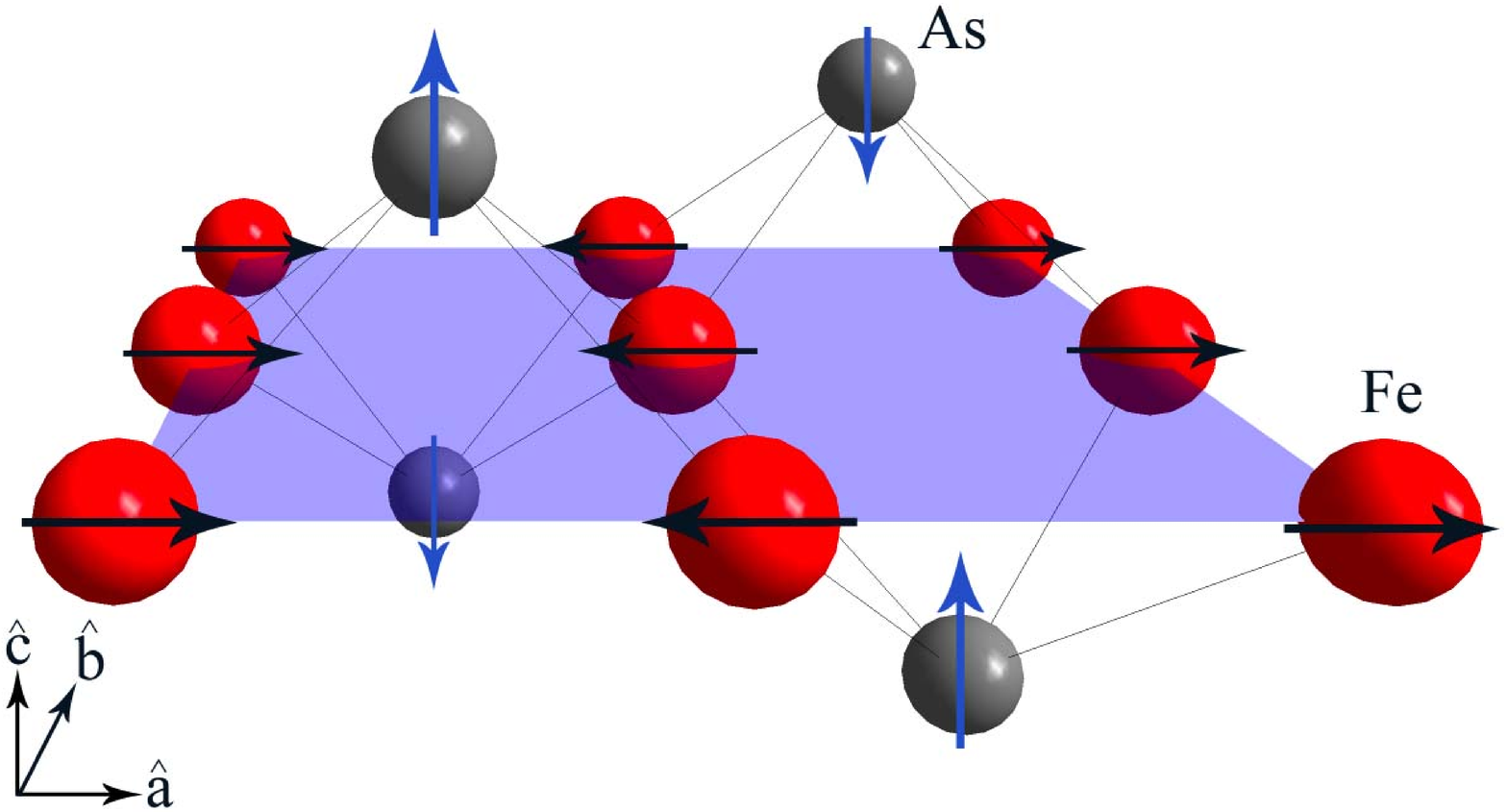}
 \caption{(Left) Zero-field spectra of \cafeas\ as a function of temperature. The dotted line represents the phase transition at $T_0=169$ K. (Right) In the ordered state, the internal magnetic field at the As sites lies in the $c$ direction. In the figure the As atoms are grey and the Fe atoms are red. The ordered Fe moments are represented by black arrows and the hyperfine fields at the As sites are blue arrows.}
\label{fig:spectrum}
\end{center}
\end{figure}

The behavior of the $I=3/2$ \as\ nuclei in \cafeas\ is determined by the nuclear spin Hamiltonian:
\begin{equation}
\label{eqn:hamiltonian}
\mathcal{H} = \gamma\hbar\hat{I}\cdot\mathbf{H}+\frac{eQV_{cc}}{4I(2I-1)}[(3\hat{I}_c^2 - \hat{I}^2) +
\eta(\hat{I}_a^2-\hat{I}_b^2)],
\end{equation}
where $\gamma$ is the gyromagnetic ratio, $\mathbf{H} = \mathbf{H}_{\rm ext}+\mathbf{H}_{\rm int}$ is the magnetic field at the nuclear site, $e$ is the electron charge, $Q$ is the quadrupolar moment,
$V_{\alpha\beta}$ are the components of the EFG tensor, and $\eta
=(|V_{aa}|-|V_{bb}|)/|V_{cc}|$ is the asymmetry parameter of the EFG.  The
NQR frequency (in zero field) is defined as: $\nu_Q =\nu_c\sqrt{1+\eta^2/3}$,
where  $\nu_c=3eQV_{cc}/2I(2I-1)h$ and $h$ is
Planck's constant. In \cafeas, the As sits in a site with axial symmetry ($\eta=0$) in the paramagnetic tetragonal phase above $T_0$, and Eq. \ref{eqn:hamiltonian} predicts a single resonance in zero external field.  Indeed, at 172K we find an NQR resonance with frequency $\nu_Q = \nu_c = 14.227(4)$ and FWHM 480kHz (Fig. \ref{fig:spectrum}), 1.8\% higher than previously reported \cite{BaekCaFe2As2PRB}.  Since the EFG at the As is a strong function of temperature and the interlayer ion, this slight difference is not surprising.
Below the first order transition at $T_0$, the structure becomes orthorhombic ($Fmmm$) and long-range antiferromagnetic order of the Fe spins sets in with an ordered moment of roughly 0.8$\mu_B$ \cite{Goldman2008PRB}.  In this case, the EFG no longer is axially symmetric at the As site and $\eta$ is non-zero.  Furthermore, a finite internal field $H_{\rm int}$ is present at the As sites as a result of the hyperfine coupling between the As nuclei and the ordered electron spins.  Therefore, even in zero external field, $H_{\rm ext}=0$, the degeneracy of the $I=3/2$ multiplet in Eq. \ref{eqn:hamiltonian} is lifted leading to three different resonances.  Fig. \ref{fig:spectrum} shows two of these resonances below $T_0$.  It is important to note that since $\eta\neq0$, the commutation relation $[\mathcal{H},\hat{I}_z]\neq 0$ and thus the eigenstates are admixtures of different $\hat{I}_z$ states.

Below $T_0$, $H_{\rm int}$, $\nu_c$ and $\eta$ all change discontinuously, and it is not straightforward to determine their values.  However, Ref. \cite{BaekCaFe2As2PRB} reports $\nu_c$ and $H_{\rm int}$ determined by field-swept spectra in an applied external field. At 4 K, we find resonances at 19.212 and 31.596 MHz, which are consistent with $\nu_c=12.38(4)$ MHz and $H_{\rm int} = 26.3(1)$ kOe along the $c-$direction.  Note that the internal field is parallel to the principal axis of the EFG tensor and that the Zeeman is larger than the quadrupolar interaction, therefore the resonance frequencies are relatively insensitive to the value of $\eta$.  In order to determine $\eta$, we would need to measure the lower resonance at roughly 7 MHz, but for technical reasons we were unable to operate in this regime.   Our results in zero external field are consistent with these previous observations measured in large external fields, suggesting that the magnetic structure is unaffected by the presence of an external field.  The advantage of zero-field NMR is that we can work with a randomly oriented powder, and therefore the signal is increased due to the larger surface area. The internal field is oriented either parallel or antiparallel to the $c-$axis and arises due to the hyperfine coupling to the ordered Fe spins as shown in the right hand side of Fig. \ref{fig:spectrum}. Neutron scattering shows an ordered moment of 0.8$\mu_B$ with an ordering wavevector $\mathbf{Q}$ along the (010) direction as shown by the black arrows. For this magnetic structure, we estimate a direct dipolar field of 3.2 kOe, roughly one order of magnitude smaller than the observed value.  In fact, the hyperfine interaction must arise from transferred spin density via the electronic wavefunctions.  By plotting the Knight shift reported in \cite{BaekCaFe2As2PRB} versus susceptibility (not shown), we find hyperfine couplings $A_c=24$ kOe/$\mu_B$ and $A_{ab}=23$ kOe/$\mu_B$, similar to the isotropic value reported in \cite{takigawa2008}.  However, since the isotropic component of the hyperfine coupling must vanish at the As site in this antiferromagnetic structure, the transferred hyperfine interaction must also contain a tensor component, probably of dipolar symmetry.

\begin{figure}
\begin{center}
 \includegraphics[width=90mm,clip]{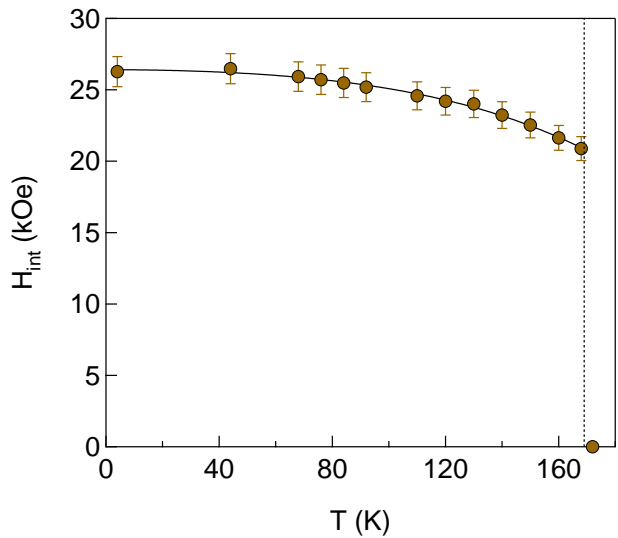}
\caption{The internal field versus temperature in the ordered state of \cafeas. The solid line is a fit to as described in the text, and the dotted vertical line indicates the phase transition at $T_0$.}
\label{fig:internalfield}
\end{center}
\end{figure}

The internal field at the As site, shown in Fig. \ref{fig:internalfield}, is directly proportional to the sublattice magnetization.  There is a clear discontinuity at $T_0$, consistent with the first order nature of the transition. In calculating $H_{\rm int}$ from the position of the resonance, we have assumed that $\nu_z$ remains temperature independent  below $T_0$, as previously reported  \cite{BaekCaFe2As2PRB}.  The solid line in the figure represents a polynomial fit the the data, which we use for fitting the \slrr\ data (see below).  Although $H_{\rm int}$ is proportional to the ordered moment and the isotropic hyperfine constants are known, the \textit{anisotropic} component has not been independently measured.  Our measurements put constraints on this tensor, which depends critically on the particular hyperfine coupling model.  Takigawa et al. have outlined a scenario based on local moments at the Fe sites; in this case, we find that $B_{ac}\sim 8$ kOe/$\mu_B$, in agreement with \cite{BaekCaFe2As2PRB}.  However, since a more itinerant description of the magnetism may be more appropriate for the ferropnictides, a different hyperfine coupling scheme may be present, involving on-site spin polarization of the electrons in the As 4s and 3p orbitals rather than transferred interactions from local moments.

\subsection*{Relaxation measurements}

\begin{figure}
\begin{center}
 \includegraphics[width=75mm,clip]{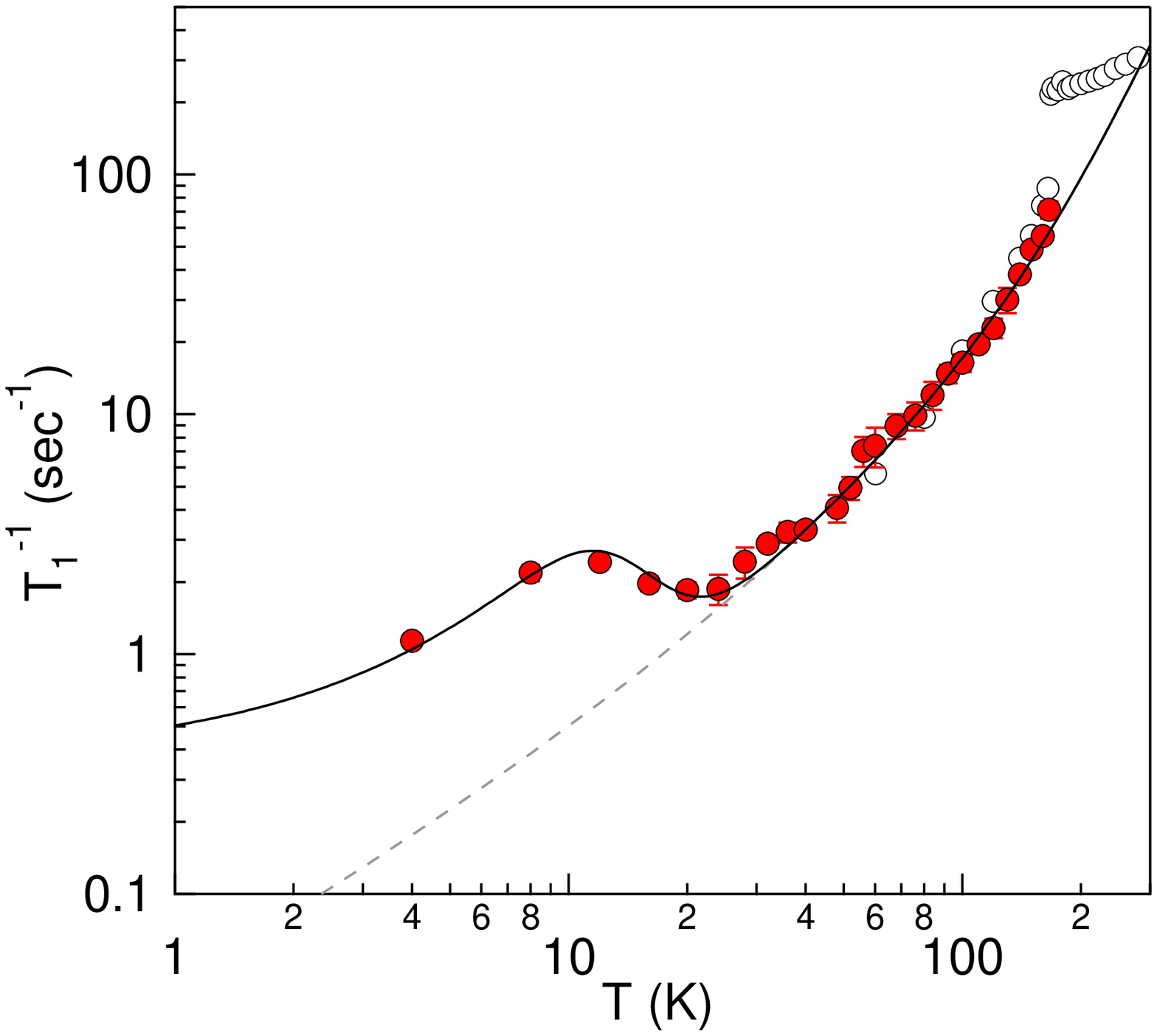}
  \includegraphics[width=75mm,clip]{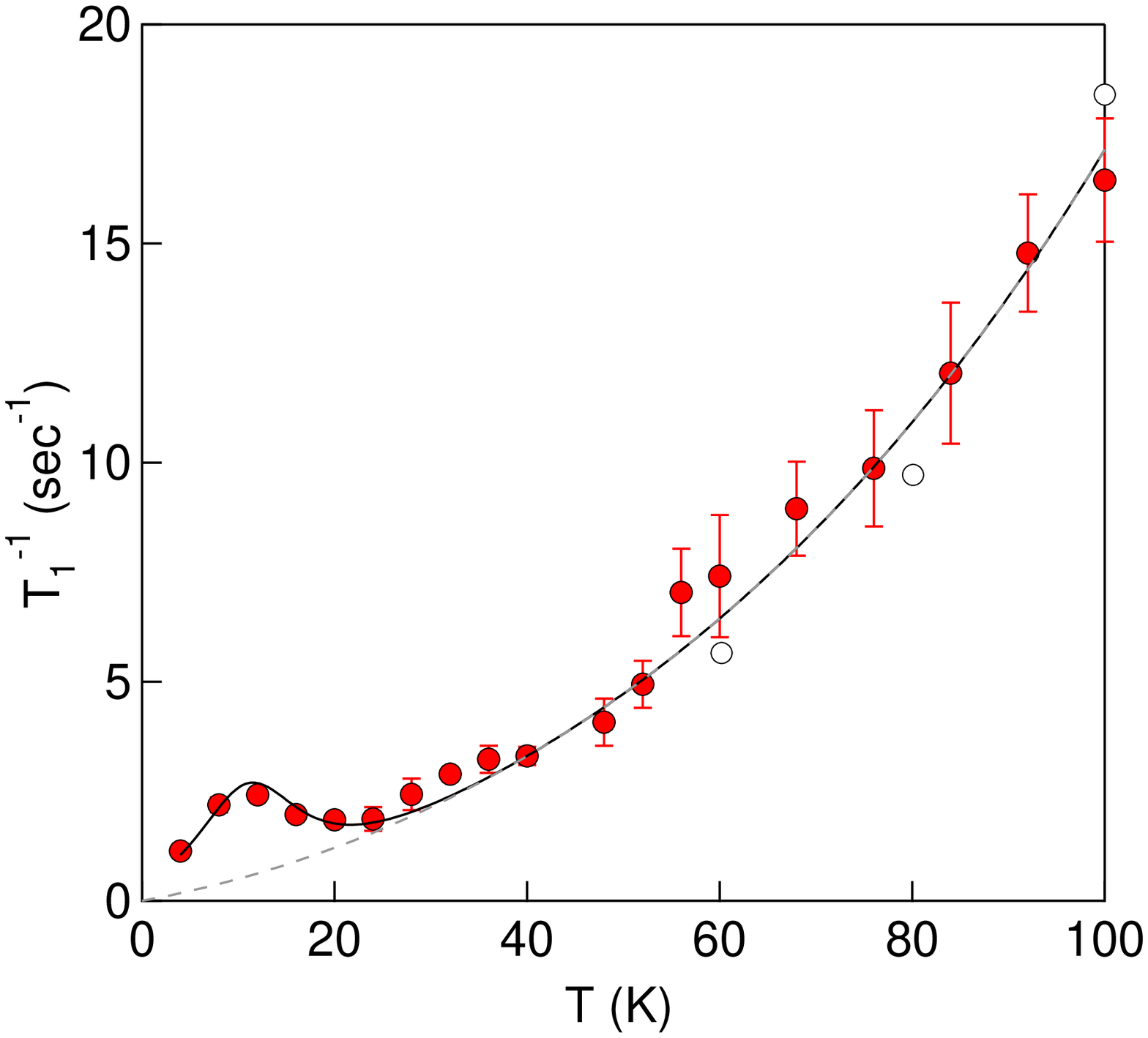}
\caption{The \slrrtext\ versus temperature in the ordered magnetic state of \cafeas, shown on a log-log scale (left) and a linear-linear scale (right).  The solid lines are a fit to the data as described in the text, and the dashed grey lines show the contribution in the absence of the glassy peak.  The small peak at 10 K may be associated with slow motion of domain walls.}
\label{fig:T1}
\end{center}
\end{figure}

Figure \ref{fig:T1} shows the \slrrtext\ versus temperature measured in the internal field at the 32 MHz transition. The data were measured by inversion recovery, and the nuclear magnetization was measured by summing up to 16 echoes in a Carr-Purcell-Meiboom-Gill (CPMG) sequence.  This procedure allowed us to significantly increase the signal to noise ratio without sacrificing acquisition time; however we found that sequences with more pulses than required for 16 echoes led to heating of the sample.  The relaxation data were fit to the standard formula for the satellite transition of a spin $3/2$ nucleus: $1-M(t)/M_0 = 0.1\exp(-t/T_1) + 0.5\exp(-3t/T_1) + 0.4\exp(-6t/T_1)$.  In principle, the relaxation form can be more complex, as certain nuclear transitions may be allowed by the fact that the eigenstates are admixtures of the $I_z$ states.  The exact form depends on the value of $\eta$, and for simplicity we have ignored this discrepancy.

In the ordered state, there are two primary mechanisms that can give rise to \slrrtext: (i) spin-flip scattering from itinerant quasiparticles,  and (ii) fluctuations of the hyperfine fields from the static ordering.  The first mechanism is the usual Korringa-type relaxation present in conductors and varies as $T_1^{-1}\sim T$.  The second mechanism arises from magnons: Goldstone excitations of the transverse degrees of freedom.  Since the ordered Fe moments point in the (100) direction, there can be fluctuations in the (010) and (001) directions. Fluctuations of both types will give rise to finite hyperfine fields at the As site, and hence will relax the nuclei.  The contribution to \slrr\ from these magnons varies roughly as $T_{1}^{-1}\sim 1-M(t)/M(0)$, where $M(t)$ is the order parameter \cite{Moriya1974,olegcomment}. Neutron scattering in SrFe$_2$As$_2$ revealed a spin gap of 6.5meV in the ordered state, so below roughly 70 K this magnon scattering mechanism should be suppressed, and the dominant relaxation mode will be Korringa relaxation \cite{INSSrFe2As22008PRL}.  Therefore we have chosen to fit the \slrr\ data to the following expression:
\begin{equation}
T_1^{-1} = aT + b(1-M(T)/M(0)) + T_{1,{\rm BPP}}^{-1},
\label{eqn:T1vsT}
\end{equation}
where $a$ and $b$ are constants, $M(T)$ is measured from the temperature dependence of the internal field, $H_{\rm int}(T)$, and $T_{1,{\rm BPP}}^{-1}$ is a third, glassy relaxation channel that gives rise to the peak at 10 K, and will be discussed below.  To fit the \slrr\ data, we first fit $H_{\rm int}(T)$ to a polynomial expression and then used the fit parameters in the expression \ref{eqn:T1vsT}. This expression fits  the relaxation data well for $0.1T_0 < T < 0.9T_0$, as seen by the dashed grey lines in Fig. \ref{fig:T1}.

The low temperature peak below 20K is unexpected, but resembles the glassy behavior observed in several lightly doped cuprate systems \cite{ChouLightlyDopedPRL,currohammel}.  We speculate that in the \cafeas, there are some low energy spin degrees of freedom with glass-like dynamics.  To model this behavior, we assume that there is a fluctuating hyperfine field $h(t)$ with autocorrelation function $\langle h(t) h(0) \rangle=h_0^2\exp(-t/\tau_c)$, where $\tau_c$ is the correlation time.  In a glassy system, $\tau_c$ typically exhibits either activated ($\tau_c=\tau_0\exp(-\Delta/T)$) or Vogel-Fulcher ($\tau_c=\tau_0\exp(-\Delta/(T-T_s))$ behavior \cite{schmalianglass}.  For concreteness, we have chosen the former in order to fit the data.  In the presence of this fluctuating field, \slrr\ will exhibit a characteristic Bloembergen-Purcell-Pound (BPP) peak when $\omega_L\tau_c=1$:
\begin{equation}
T_{1,{\rm BPP}}^{-1}(T) = \frac{\gamma^2 h_0^2 \tau_c(T)}{1+\omega_L^2\tau_c^2(T)},
\label{eqn:T1BPP}
\end{equation}
where $\omega_L$ is the Larmor frequency.  As seen by the solid lines in Fig. \ref{fig:T1}, the data are  fit well by this expression for $\Delta = 4.5$ K.  It is interesting to note that this value is close to interlayer magnetic exchange interaction estimated by LDA calculations \cite{MazinNature2009}.

It is possible that the origin of these slow glass-like spin dynamics could be extrinsic impurities in our samples that give rise to quenched disorder.  However, the fact that the bulk properties are identical to the values in the literature, particularly the sharp first-order transition at 169 K, suggest otherwise \cite{BaekBaFe2As2PRB2008}.  Therefore, we speculate that these slow spin dynamics may be an intrinsic phenomena associated with the motion of domain walls in the ordered state.  The columnar stripe-like antiferromagnetic order in this system can support several types of domains and defect structures, including both anti-phase boundaries, stacking faults along the $c-$direction and twin boundaries \cite{MazinNature2009}.  Recently Mazin and Johannes argued that since the magnetic phase transition is first order, it is reasonable to expect that different domains emerge naturally, with mobile boundaries giving rise to low energy fluctuations (long time scales) and possibly reduced effective moments.  These mobile domain boundaries may be the source of the low temperature peak we observe in \slrr.

A similar situation exists in the antiferromagnetic state of the doped cuprate oxides but with important differences.  $\mu$SR and NMR data in the antiferromagnetic state of \lsxco\ show an ordered moment that is reduced from the full value of the Cu 3d$^9$ $S=1/2$ spin moments \cite{BorsaLSCOorderparameterdopingPRB,Klauss2004}.  At $T\sim 10$ K, the \slrrtext\ exhibits a dramatic peak associated with slow glassy spin fluctuations, and below this temperature the ordered moment probed by the muon or nuclear spins increases dramatically to its full $S=1/2$ value.  The interpretation is that there are collective motions of the doped holes in the checkerboard antiferromagnetic background that are an intrinsic behavior of the correlated system \cite{schmalianglass}.  As these collective motions freeze below the motionally narrowed limit, the linewidth broadens to reflect the full ordered moment.

Such an increase in the ordered moment below the freezing temperature is not observed in our data.  Actually the internal field remains constant down to 2K, suggesting that the ordered moments are not reduced by the presence of domain wall motions in this system.  It is possible that further experiments below 1K may detect such a change, but a more likely explanation is that the ordered moments are reduced either by spin-orbit and hybridization effects, or by proximity to a quantum phase transition \cite{Uhrig2008}.  The columnar antiferromagnetic structure of the ferropnictides is a natural consequence of a Heisenberg model involving local spins and both nearest neighbor $(J_1$) and next-nearest neighbor ($J_2$) couplings.  Although it remains unclear whether or not the magnetic order of the ferropnictides can be described by such a localized model, the dispersion measured by inelastic neutron scattering is well described by this model \cite{INSSrFe2As22008PRL,INSBaFe2As2PRB2008}.  Recently, Uhrig and coworkers have pointed out that this model undergoes a quantum phase transition for sufficiently large $J_1/J_2$ from a stripe-like ground state to a checkerboard state. In this case, the ordered moments are reduced, yet multiple types of domains can exist.  It is possible, then, that domain wall motion can give rise to a BPP peak in \slrr\ yet the magnitude of the ordered moment is not reduced by domain wall motion but rather quantum critical effects.

\section*{Conclusions}

We have measured the spectra and spin lattice relaxation rate in the antiferromagnetically ordered state of \cafeas.  The origin of the small ordered moment in these materials has remained mysterious despite nearly one year of intense research. Our \slrrtext\ data reveal the presence of both quasiparticle (Korringa) scattering and magnon excitations.  Surprisingly, \slrr\ exhibits a peak at 10K, which is well characterized by an activated spin fluctuation process with energy $\sim 5$K.  We speculate that these slow dynamics are related to the freezing of domain walls in this system.  Mobile domain walls may be a natural consequence of the columnar stripe antiferromagnetism that sets in below a first-order transition.  It is crucial to investigate the evolution of these low energy glassy dynamics as a function of doping in order to understand the disappearance (or coexistence) of antiferromagnetism and superconductivity.

\section*{Acknowledgements}

We thank S.-H. Baek, R. Singh, and O. Sushkov for numerous enlightening discussions.

\section*{References}

\end{document}